\documentclass[10pt]{iopart}
\usepackage{iopams}
\usepackage[dvips]{graphicx}
\usepackage[scriptsize,bf]{caption}
\usepackage{xspace}
\eqnobysec
\newcommand{\const}{\textit{const.}}

\renewcommand{\br}[1]{\left(#1\right)}
\newcommand{\ud}[1]{\mathrm{d}#1}
\renewcommand{\exp}[1]{e^{#1}}
\newcommand{\sq}[1]{\left[#1\right]}

  \newcommand{\pp}{\mathcal{P}}
\newcommand{\qq}{\mathcal{Q}}
\newcommand{\hh}{\mathcal{H}}
\renewcommand{\ll}{\mathcal{L}}
\newcommand{\unit}{\mathbf{E}}
\newcommand{\nnt}{\mathbf{NN^T}}
\newcommand{\nvt}{\mathbf{NV^T}}
\newcommand{\nomt}{\mathbf{N\Omega^T}}
\newcommand{\vnt}{\mathbf{VN^T}}
\newcommand{\vvt}{\mathbf{VV^T}}
\newcommand{\vot}{\mathbf{V\Omega^T}}
\newcommand{\ont}{\mathbf{\Omega{}N^T}}
\newcommand{\ovt}{\mathbf{\Omega{}V^T}}
\newcommand{\oot}{\mathbf{\Omega{}\Omega^T}}
\newcommand{\tA}{\left(\pp{F_{,\,\pp\qq}}+2\br{{F_{,\,\qq}}+\qq{F_{,\,\qq\qq}}}\right)}
\newcommand{\tB}{\left({F_{,\,\pp}}+\pp{F_{,\,\pp\pp}}+2\,\qq{F_{,\,\pp\qq}}\right)}
\newcommand{\tC}{\left(F-{\pp{}F_{,\,\pp}}\right)}
\renewcommand{\sp}[2]{\mathbf{{#1}^T{#2}}}
\newcommand{\eqref}[1]{\eref{#1}}
\newcommand{\tfrac}[2]{\case{#1}{#2}}
\begin{document}
\title[Breathing Relativistic Rotators and Fundamental Dynamical Systems]{Breathing Relativistic Rotators\\ and Fundamental Dynamical Systems}
\author{{\L}ukasz Bratek}
\address{Henryk Niewodnicza{\'n}ski Institute of Nuclear Physics, \\
Polish Academy of Sciences, Radzikowskego 152, PL-31342 Krak{\'o}w, Poland}
\ead{lukasz.bratek@ifj.edu.pl}

\begin{abstract}
 Recently, it was shown, that the mechanical model of a massive spinning particle proposed by Kuzenko, Lyakhovich and Segal in 1994, which is also the fundamental relativistic rotator rediscovered independently 15 years later by Staruszkiewicz in quite a different context, is defective as a dynamical system, that is, its Cauchy problem is not well posed. This dynamical system is fundamental, since its mass and spin are parameters, not arbitrary constants of motion, which is a classical counterpart of quantum irreducibility.
  It is therefore desirable to find other objects which, apart from being fundamental, would also have well posed Cauchy problem.

 For that purpose, a class of breathing rotators is considered. A breathing rotator consists of a single null vector associated with position and moves in accordance with some relativistic laws of motion. Surprisingly,  breathing rotators which are fundamental, are also defective as  dynamical systems. More generally, it has been shown, that the necessary condition for a breathing rotator to be similarly defective, is functional dependence of its Casimir invariants of the Poincar{\'e} group.
 \end{abstract}
\pacs{03.30.+p, 45.50.-j}
\submitto{}
\maketitle

\section{Introduction}

This work investigates a class of breathing rotators which are relativistic dynamical systems consisting of a single null vector $k$ associated with position $x$.

To reduce, with a single clear-cut physical idea, the enormous variety of  relativistically invariant  actions possible for a dynamical system,  one can use the notion of a fundamental dynamical system defined by Staruszkiewicz \cite{bib:astar1}. It can be rephrased as follows:
\begin{quote}
 {\textit{A dynamical system described by a relativistically invariant action, is said to be  fundamental, if its both Casimir invariants of the Poincar\'{e} group are parameters with fixed numerical values rather than  arbitrary constants of motion}}.\end{quote}
 By applying this definition, one obtains two independent constraints that must be satisfied by the action of a dynamical system. These two constraints are referred to here as \textit{fundamental conditions}. It is clear, that Casimir invariants of other symmetry groups of the action, could be fixed in a similar way.

Representations of relativistic quantum mechanical systems are labeled by numerical values of
Casimir invariants of different symmetry groups, among which the most important is the Poincar\'{e} group \cite{bib:wigner}. Fundamental conditions are classical counterpart of quantum irreducibility.

It is obvious that this \textit{idea of fundamentality} already suffices to fully justify the necessity of research on relativistic fundamental classical systems. Another reason is that a subset of fundamental systems, whose motion is periodic in their centra of mass frame, could be used as ideal classical clocks.  These purely mathematical constructs, with their own intrinsic non-quantum clocking mechanisms, experiencing no fatigue and friction, are a way to study some difficult and not well understood
problems in special and general theory of relativity \cite{bib:astar1}.

   As an example of a fundamental dynamical system can be considered a simple geometrical model for a relativistic massive point-like particle of arbitrary spin, constructed in 1994 by Kuzenko, Lyakhovich, and Segal \cite{bib:segal} in the context of some issues in the dynamics of spinning particles  in high-energy physics. It is a Poincar{\'e}-invariant dynamical system in Minkowski space-time, whose spin sector is represented by two additional, intrinsic degrees of freedom on the unit sphere. The Lagrangian for this particle was found uniquely by fixing the values of mass and spin, which is essentially the realization of the above idea of fundamentality.

   Fifteen years later, in quite a different context, aiming to construct an ideal mathematical clock, Staruszkiewicz rediscovered, completely independently, the same mathematical entity, identified this time as the fundamental relativistic rotator \cite{bib:astar1} -- a particular realization of a rigid body of Hanson and Regge \cite{bib:hanson} with the number of null directions reduced to one (a null direction is a class of all collinear null vectors). A rigid body of Hanson and Regge, or a spherical top, consists of a point on a world-line in Minkowski space-time and a rotating frame attached to that point \cite{bib:hanson}. However, it can be equivalently characterized as
a dynamical system described by position and three null directions \cite{bib:astar1}. This observation led Staruszkiewicz to his definition of a relativistic rotator
 with five degrees of freedom (which is not a rigid body of Euler which has six degrees of freedom) \cite{bib:astar1}.
     The Lagrangian for this rotator was determined uniquely by requiring that the rotator should be a fundamental dynamical system. 
     
     Mathematical equivalence of the point-like spinning particle of Kuzenko, Lyakhovich, and Segal  and the fundamental relativistic rotator of Staruszkiewicz, is seen directly from one-to-one correspondence between the points on the surface of the unit sphere and null directions in Minkowski space-time. 
   It is remarkable, that this dynamical system is uniquely determined by the fundamental conditions.

The fundamental relativistic rotator
 can indeed be used as an ideal clock when its motion is periodic in its center of mass frame. Such a clock is perfect in the sense, that the values of its mass and intrinsic spin are unaffected even when it interacts with external fields. This interesting property was confirmed by Kassandrov, Markova, Schaefer, and Wipf \cite{bib:schaefer}, who added to the Lagrangian of the fundamental relativistic rotator the usual interaction term with electromagnetic field, as if the rotator was a structureless charged particle.

Surprisingly, it has been shown recently in  \cite{bib:bratek}, that the fundamental relativistic rotator, or equivalently, the massive spinning particle, is defective as a dynamical system  -- accelerations are not uniquely determined from velocities and positions.
This defect disappears by arbitrary small deformation of the Lagrangian in the class of Lagrangians with the same symmetries, however, on the cost of losing the central feature of being fundamental. This shows that the defectiveness is not caused by the number of degrees of freedom but is inherent to the fundamental relativistic rotator. It will also be evident from the analysis presented later, that the interaction term proposed in \cite{bib:schaefer} cannot remove this degeneracy, because it was assumed linear in velocities.

 One can hypothesize that a fundamental dynamical system must be sufficiently complex in order to ensure well-posedness of the Cauchy problem.
In this respect, by considering breathing rotators, it is examined if inclusion of an additional degree of freedom  would suffice to  eliminate this degeneracy, not violating fundamental conditions at the same time. Unfortunately, as will be shown later, breathing rotators are  still too simple dynamical systems to be fundamental.

Notation. Metric tensor is such that $\br{\ud{s}}^2=\br{\ud{x^0}}^2-\br{\ud{x^1}}^2-\br{\ud{x^2}}^2-\br{\ud{x^3}}^2$. Scalar product of four-vectors $a^{\mu}$ and $b^{\mu}$ is denoted by $ab$. Partial derivatives w.r.t scalars $p$, $q$ are denoted by a coma sign: e.g. $F,_{p}\equiv\tfrac{\partial\,F}{\partial\,p}$, $F,_{pq}\equiv\tfrac{\partial^2F}{\partial{}q\partial{}{p}}$.
\section{Hessian determinant and functional dependence of Casimir invariants}

There are four nonzero Poincar\'{e} invariants formed from $x$,  $k$ and their first derivatives: $\dot{x}\dot{x}$, $k\dot{x}$, $\dot{k}\dot{x}$, $\dot{k}\dot{k}$. Here, a dot sign denotes differentiation with respect to arbitrary parameter along a world-line.
However, not every combination of the invariants is suitable for a relativistically invariant Hamilton's action. Such an action must be reparametrization invariant. In addition, in order to reduce the number of free parameters to minimum, this action is assumed to be independent of the physical dimension of null vector $k$.
Consequently, satisfying these requirements, the most general action for breathing rotators reads
\begin{equation}
\fl -m\int\sqrt{\dot{x}\dot{x}}\,F\br{{\pp},{\qq}}\ud{\tau},
\qquad {\pp}=\ell\frac{\dot{k}\dot{x}}{k\dot{x}\sqrt{\dot{x}\dot{x}}}, \quad  {\qq}=-\ell^2\frac{\dot{k}\dot{k}}{\br{k\dot{x}}^2}.\label{eq:action}\end{equation}
Dimensional constants $m$ (mass) and $\ell$ (length) are the only parameters of the model.
The momenta canonically conjugated with $x$ and $k$ are $P_{\mu}\equiv-\frac{\partial{}L}{\partial{}\dot{x}^{\mu}}$ and  $\Pi_{\mu}\equiv-\frac{\partial{}L}{\partial{}\dot{k}^{\mu}}$, or explicitly,
  \begin{eqnarray*}P=m\,
  \br{ \left( F - {\pp} \,F_{,\pp}
     \right)\,u
  - \left( 2\,{\qq} \,F_{,\qq} +
       {\pp} \,F_{,\pp} \right)\frac{k\, }{{ku}}
       +{\pp} \,F_{,\pp}\,
  \frac{\,\dot{k}}{{\dot{k}u}} },\\
  \Pi=\frac{m\,\ell}{ku}
  {\left(  F_{,\pp}\,u -2\,{\pp} \,F_{,\qq}\frac{\,\dot{k}\,}{{u\dot{k}}}
      \right) }, \end{eqnarray*} where $u\equiv{\dot{x}}/{\sqrt{\dot{x}\dot{x}}}$.

As a consequence of relativistic invariance of action \eqref{eq:action}, the identity $0=\int_{\tau_1}^{\tau_2}\delta{L}=
-\br{P\delta{x}+\Pi\delta{{k}}}_{\tau_2}+\br{P\delta{x}+\Pi\delta{{k}}}_{\tau_1}
 $ holds for infinitesimal Poincar\'{e} transformations of solutions (for the purpose of this section it suffices  to keep in mind that $kk=0$, however, in order to find equations of motion in a covariant form, one must add to the Hamilton's action an appropriate term with a Lagrange multiplier). The invariance with respect to space-time translations,
$\delta{}x^{\mu}=\epsilon^{\mu}=\const$ and $\delta{k}^{\mu}=0$, implies conservation of momentum $P^{\mu}$, while the invariance with respect to space-time rotations, $\delta{x}^{\mu}=\Omega^{\mu}_{\phantom{\mu}\nu}x^{\nu}$ and $\delta{k}^{\mu}=\Omega^{\mu}_{\phantom{\mu}\nu}k^{\nu}$ ($\Omega_{\mu\nu}=\const$ and $\Omega_{(\mu\nu)}=0$), implies conservation of angular momentum $M_{\mu\nu}\equiv{}x_{\mu}P_{\nu}-x_{\nu}P_{\mu}+k_{\mu}\Pi_{\nu}-k_{\nu}
\Pi_{\mu}$. These constants of motion are used to form both of the Casimir invariants of the Poincar\'{e} group: $P_{\mu}P^{\mu}$ and $W_{\mu}W^{\mu}$, where
 $W^{\mu}$ is the Pauli-Luba\'{n}ski (space-like) spin-pseudovector: $W^{\mu}:=-\frac{1}{2}\epsilon^{\mu\alpha\beta\gamma}M_{\alpha\beta}P_{\gamma}$, hence
\begin{eqnarray*}\fl \phantom{WWW}
     PP&=&m^2\,\left( \left( F - {}{\pp\,F_{,\pp}}\,  \right) \,
     \left( F - {}{\pp\,F_{,\pp}}\,  -
       4\,{}{\qq \,F_{,\qq}}\, \right)  -
    {{\qq\,}{F_{,\pp}}}^2 \right),\\
    \fl \phantom{WWW}
           WW&=&-G(P,\Pi,k)=
-m^4\,{\ell}^2 \qq\,{\left( {{}{F_{,\pp}}}^2 +
      2\,{}{F_{,\qq}}\,\left( F - {\pp\,}{F_{,\pp}}\,
         \right)  \right) }^2,\end{eqnarray*}
where $G(P,\Pi,k)$ is the determinant of Gramian matrix of scalar products of four-vectors $P$, $\Pi$ and $k$.

Function $F$ should not be chosen at random, but be determined uniquely based on some clear-cut physical idea. In this respect fundamental conditions are imposed. The unspecified arbitrary parameters $m$ and $\ell$ can now be set by relating them directly to the fixed numerical values of the Casimir invariants. With no loss to generality, this can be done by requiring that
\begin{equation} PP\equiv{}m^2, \qquad WW\equiv-\frac{1}{4}m^4\ell^2.\label{eq:fundcond}\end{equation}
This gives two completely unrelated differential equations that must be simultaneously satisfied by function $F$.  It is clear that there is no a priori reason for the existence of such a common solution. Remarkably enough, two such solutions are possible  (c.f. \ref{app:SolvCond}), giving rise to two fundamental breathing rotators, discussed in more detail in the Conclusions.

Unfortunately, similarly as the fundamental relativistic rotator, fundamental breathing rotators turn out defective as dynamical systems due to vanishing of the Hessian determinant. Here, by a Hessian determinant is understood the determinant of a matrix  of second derivatives of a Lagrangian with respect to velocities associated only with the dynamical degrees of freedom.

Best to see this quickly is the following, astounding relationship of the Hessian determinant (denoted by $\det{\mathcal{H}}$) with a Jacobian determinant of the following $F$-dependent mapping, leading from coordinates  $\br{\pp,\qq}$ to coordinates $\br{PP,WW}$  (c.f. \ref{app:hessian} for a derivation)
\begin{equation}\label{eq:HessianJacobian}\det{\mathcal{H}}=\kappa\cdot
\frac{F-\pp\,F_{,\pp}}{F_{,\pp}\br{\pp^2+\qq}-\pp{}F}
\cdot\,\left|\frac{\partial\br{PP,WW}}{\partial\br{\pp,\qq}}\right|.\end{equation}
Here, $\kappa$ is some kinematical factor, the same for all $F$.
In the distinguished case, when $F=\sqrt{1+\frac{\pp^2}{\qq}}S\br{\qq}$ (then $F_{,\pp}\br{\pp^2+\qq}-\pp{}F=0$),
the Jacobian determinant vanishes, but not necessarily does the Hessian determinant (indeterminate form $\tfrac{0}{0}$).
Indeed, then Casimir invariants are functionally dependent: $PP=m^2S\br{S-4\qq{}S'}$ and $WW=-\br{2m^2\ell{}S\sqrt{\qq}S'}^2$,  while $\det{\mathcal{H}}\propto\frac{\qq{}S^3S'}{\br{\pp^2+\qq}^2}
\br{2\qq\br{S'}^2+S\br{S'+2\qq{}S''}}$, that is, $\det{\mathcal{H}}\propto{}
S^3S'\br{PP}'\propto{}S^2\br{WW}'$, which is nonzero unless fundamental conditions are imposed.
In all other cases, when $F_{,\pp}\br{\pp^2+\qq}-\pp{}F\neq 0$,  vanishing of the Hessian determinant is equivalent to vanishing of the Jacobian determinant (if $F-\pp{}F_{,\pp}=0$ then $WW=m^2{\ell^2}PP$, which is unphysical  as then $PP<0$ because pseudovector $W$, being orthogonal to a null vector, is always space-like).

\section{Conclusions}
In this paper a class of relativistic dynamical systems, described by a single null vector associated with a space-time position and defined by a relativistically invariant action \eqref{eq:action}, has been examined. To distinguish them from the class of rotators considered by Staruszkiewicz \cite{bib:astar1}, the systems are called "breathing rotators".  Breathing rotators have six dynamical degrees of freedom -- three for position, two for the null direction associated with null vector $k$, and one "breathing" degree of freedom associated with the amplitude of $k$. As usual in relativity theory, Lagrangians must be reparametrization invariant, thus the arbitrary parameter $\tau$ is not dynamical and is treated as  a gauge variable.

There are two subclasses of breathing rotators that are distinguished by analytical properties of relation \eqref{eq:HessianJacobian}. Rotators with singular Hessian form a subset of breathing rotators with functionally dependent Casimir invariants $PP$ and $WW$, whereas functional independence of the invariants guarantees  non-singularity of the Hessian.

Another result is that breathing rotators that are fundamental have singular Hessian. This property makes them defective as dynamical systems. So far, this has been the second example  of relativistically invariant systems known, for which  fundamental conditions imply singularity of the Hessian. The other is the massive spinning particle \cite{bib:segal}, or equivalently, the fundamental relativistic rotator \cite{bib:astar1}, for which this property has been discovered in \cite{bib:bratek}.  However, it seems rather improbable that fundamental conditions would always imply singularity of the Hessian. Also, a proof of such a theorem seems hopelessly difficult in a generic situation. Therefore it is necessary in the future to construct a counterexample.

As follows from \ref{app:SolvCond}, two breathing fundamental rotators are possible with the following Hamilton's actions:
\begin{equation}
\label{eq:StarlikeAction}\fl S=-m\int\ud{\tau}\sqrt{\dot{x}\dot{x}}\sqrt{
 \sq{1-\frac{(\dot{k}\dot{x})(\dot{k}\dot{x})}
 {(\dot{x}\dot{x})(\dot{k}\dot{k})}}
 \sq{1\pm\sqrt{-\ell^2\frac{\dot{k}\dot{k}}{\br{k\dot{x}}^2}}}},
 \end{equation}
 \begin{equation}
\label{eq:myaction}
\fl
S_{\nu}=-m\int\ud{\tau}\sqrt{\dot{x}\dot{x}}
\br{
 \sqrt{1\pm\sqrt{-\ell^2\frac{\dot{k}\dot{k}}{\br{k\dot{x}}^2}}+
 \nu^2\,{\ell^2}\frac{\dot{k}\dot{k}}{\br{k\dot{x}}^2}  }+\nu\,{\ell}\frac{\dot{k}\dot{x}}{k\dot{x}\sqrt{\dot{x}\dot{x}}} },\quad \nu\in\mathbb{R}.\end{equation}
  Parameter $\nu$ is an integration constant of fundamental conditions, and can be reinterpreted as an additional length scale: $\nu\ell$. It should be stressed here, that Casimir invariants for action \eqref{eq:myaction} are independent of $\nu$. For both rotators $$PP=m^2,\qquad WW=-\frac{1}{4}m^4\ell^2.$$
    Contrary to rotator \eqref{eq:StarlikeAction} which has six degrees of freedom,
    rotator \eqref{eq:myaction} must be treated as having only five degrees of freedom, since the amplitude of $k$ in this case is a gauge variable.
 Indeed, for any function $\psi(\tau)$ $$S_{\nu}[x,\exp{\psi}k]=S_{\nu}[x,k]-m\,\ell\,\nu\,\psi(\tau).$$  Since the corresponding Lagrangians differ by a total derivative, the form of equations of motion is left unchanged.  This means that the breathing mode separates completely from the dynamics of the other degrees of freedom and does not influence them at all, therefore it can be completely ignored.  As a result, the dynamical system defined by action \eqref{eq:myaction} depends on position and a null direction only, similarly as fundamental relativistic rotator.
Unfortunately, rotator \eqref{eq:myaction} cannot be considered as a replacement for the fundamental relativistic rotator (or the massive spinning particle), which is obtained by setting $\nu=0$,
$$S_{\nu=0}=-m\int\ud{\tau}\sqrt{\dot{x}\dot{x}}
\sqrt{1+\sqrt{-\ell^2\frac{\dot{k}\dot{k}}{\br{k\dot{x}}^2}}},
$$
since the determinant of a reduced $5\times5$ Hessian matrix  corresponding to the five dynamical degrees of freedom  of rotator \eqref{eq:myaction} (the breathing mode is excluded), vanishes as well. Summing up, these are  the fundamental conditions that are responsible for the singular behavior of both breathing rotators \eqref{eq:StarlikeAction} and \eqref{eq:myaction}.

\appendix
\section{\label{app:SolvCond}Solution of fundamental conditions}

To solve fundamental conditions \eqref{eq:fundcond}, it is convenient to recast them into the equivalent form
 \begin{eqnarray*}
\left. \begin{array}{r} 4\,u^2-4\,u\,\left( 1 + x\,{u_{,x}} + y\,{u_{,y}} \right) + 2\,x\,y\,{u_{,x}}\,{u_{,y}} + \left(  y^2 -x^2  \right) \,{{u_{,y}}}^2 =
  0\\
 2\,u  +  2\,u\,{u_{,x}}\, - y\,{u_{,x}}\,{u_{,y}}  + x\,{{u_{,y}}}^2= 0
  \end{array}\right\},
 \end{eqnarray*}
where $x\equiv\pm\sqrt{\qq}$, $y\equiv{}\pp$ and  $\pm\sqrt{u\br{x,y}}\equiv{}F(\pp,\qq)$, $\quad u>0$.
The second equation can be linearized by the Legendre transformation
 $u(x,y)\to x\xi+y\eta-\omega(\xi,\eta)$, $ x\to\omega_{,\xi}$, $ y\to\omega_{,\eta}$, $ u_{,x}\to\xi$, $ u_{,y}\to\eta$ going over into
 $\eta \,\left( 2 + \xi  \right) \,{{\omega }_{,\eta }} +
   \left( {\eta }^2 + 2\,\xi \,\left( 1 + \xi  \right)  \right) \,{{\omega }_{,\xi }} =
  2\,\left( 1 + \xi  \right) \,\omega$. The inhomogeneous term is removed by substitution $\omega (\xi ,\eta ) ={\left( {\eta }^2 + {\xi }^2 \right) \,h(\xi ,\eta ) }/\br{{\eta }^2 - 2\,\xi }$, giving the equivalent equation $\eta \,\left( 2 + \xi  \right) \,{h_{,\eta }} +
   \left( {\eta }^2 + 2\,\xi \,\left( 1 + \xi  \right)  \right) \,{h_{,\xi }} = 0$ which is solved by noting that gradient $\{h_{,\xi },h_{,\eta }\}$ must be collinear with vector $\{ \eta \,\left( 2 + \xi  \right) ,-{\eta }^2 - 2\,\xi \,\left( 1 + \xi  \right) \}$. The latter is proportional to a gradient of any function of a single argument $s$, $s\equiv{{\sqrt{{\eta }^2 + {\xi }^2}}}/\br{{\eta }^2 - 2\,\xi }$. Hence, the general solution for $\omega$ must be of the form $\omega(\xi ,\eta )  = s\,g(s)\sqrt{{\eta }^2 + {\xi }^2} $ with arbitrary function $g$.

    By means of the same Legendre transformation and with the obtained ansatz for $\omega$, the first equation goes over into
     $${\eta }^2\,s^2\sq{\,4\left( 1 - s^2 \right) {g}^2 +
     4\,\left( 1 - 2\,s^2 \right)s \,g' g   + \left( 1 - 4s^2 \right)s^2 g'^2 +4\,g+
     4 s\, g' \, }  = 0.$$
     It is remarkable, that all coefficients in the square bracket could be expressed by $s$ alone, showing that two  quite distinct equations of different physical origin, namely, the fundamental conditions, have a common solution. The terms linear in $g$ and $g'$ can be absorbed by introduction of function $g(s)\equiv{}G(s)-\br{2s^2}^{-1}$, and next, the term proportional to $GG'$ can be absorbed with the help of function $f(s)$ defined as $s^2\,G(s)\equiv\sqrt{1-4s^2}\,f(s)$. One is now left only with the simple equation $1 = 4\,{f}^2-{\left( 1 - 4\,s^2 \right) }^2\,\br{f'}^2 $. Its first, trivial solution, is $f(s)=\pm1/2$. The other solution can be found by substitution $f(s)=\pm\cosh\br{\psi(s)}/2$, hence $\br{\br{1-4s^2}\psi'}^2=4$, which is easily integrable and has an integration constant $\alpha$. Finally, the two resulting solutions for $g$ are: $g(s) = \br{-1 \pm  {\sqrt{1 - 4\,s^2}}}/\br{2\,s^2}$ and $g(s) = \br{-1  \pm \,\cosh (\alpha ) + 2\,s\,\sinh (\alpha )}/\br{2\,s^2}$. The corresponding solutions for $\omega$ are $
   \omega (\xi ,\eta )  = \xi-{\eta \,\left( \eta \pm \,{\sqrt{{\eta }^2 - 4\,\left( 1 + \xi  \right) }} \right) }/2$ and $\omega (\xi ,\eta ) ={\left( {\eta }^2 - 2\,\xi  \right) \,\left( -1 \pm \,\cosh (\alpha ) \right) }/
    {2} + {\sqrt{{\eta }^2 + {\xi }^2}}\,\sinh (\alpha )$, respectively, the latter with a real parameter $\alpha$. The last point is to apply the inverse Legendre transformation to obtain the corresponding solutions for $u(x,y)$ and thence for $F(\pp,\qq)$.

    Finally, there are two solutions of fundamental conditions \eqref{eq:fundcond}
\begin{eqnarray*}
F(\pp,\qq)=\pm\sqrt{
\br{1\pm\sqrt{\qq}}
\br{1+\frac{\pp^2}{\qq}}
},\\
F(\pp,\qq)=\frac{1}{a}\br{\pp\pm
\sqrt{\br{1\pm\sqrt{\qq}}a^2-\qq}},\quad a\in\mathbb{R}, \end{eqnarray*} where $a=2\sinh\br{\alpha/2}$ (repeated $\pm$ signs in a solution are not related to each other). A formal limit $a\to\infty$ reproduces the Lagrangian of the fundamental relativistic rotator.

\section{\label{app:hessian}Hessian determinant for breathing rotators}

Consider determinant of a matrix of second derivatives of the Lagrangian in action \eqref{eq:action} with respect to generalized velocities, associated with the physical degrees of freedom only. Prior to calculation of it, a convenient map of internal coordinates is to be chosen, and any spurious degrees of freedom eliminated (fixing 'gauge'). The result is always a factor of the same invariant -- a second order differential operator with respect to Lorentz invariants $\pp$ and $\qq$ and acting on function $F$ -- times an unimportant geometrical factor dependent on the particular map chosen. For the purpose of this paper, this differential invariant (up to a constant factor) is called the Hessian determinant.

To calculate it, one must fix the arbitrary parameter $\tau$. Let it be the time coordinate in a given inertial coordinate frame, times a constant dimensional factor, $\tau\equiv{}\ell^{-1}x^0$. One may chose a map of internal coordinates in which $\dot{x}=\ell\sq{1,\mathbf{V^T}}$, $k=\exp{\Psi}\sq{1,\mathbf{N^T}}$ with $\sp{N}{N}=1$, $\dot{k}=\exp{\Psi}\sq{\sp{N}{\Omega},\mathbf{\Omega^T}}$, and $\mathbf{\Omega}\equiv{}\dot{\mathbf{N}}+\dot{\Psi}\mathbf{N}$ ($\dot{\Psi}\equiv{}\sp{N}{\Omega}$), where $\mathbf{V}$ and $\mathbf{\Omega}$ stand for the $6$ generalized velocities associated with position $x$ and null vector $k$. Here, bold capitals stand for column 3-vectors, then $\sp{X}{Y}\equiv\mathbf{Y^TX}$ is the scalar product of vectors $\mathbf{X}$ and $\mathbf{Y}$, while $\mathbf{XY^T}$ is a $3\times3$ matrix (in general distinct from  $\mathbf{YX^T}$).  Up to a constant factor, the Lagrangian in action \eqref{eq:action} is given by a dimensionless scalar $\ll$
\begin{small}
\begin{equation*}\fl
\ll=-\sqrt{1-\sp{V}{V}}F\br{\pp,\qq},\qquad \pp=\frac{1}{\sqrt{1-\sp{V}{V}}}\frac{\sp{N}{\Omega}-\sp{V}{\Omega}}{{1-\sp{N}{V}}},\quad \qq=\frac{\sp{\Omega}{\Omega}-\br{\sp{N}{\Omega}}^2}{\br{1-\sp{N}{V}}^2}.\end{equation*} \end{small}
The $6\times6$ square matrix $\hh$ of second derivatives of $\ll$ with respect to 3-velocities represented by column vectors $\mathbf{V}$ and $\mathbf{\Omega}$,  has a block structure
$$\hh=\left[\begin{array}{cc}
\ll_{\vvt}&\ll_{\vot}\\ \br{\ll_{\vot}}^T & \ll_{\oot}\end{array}\right],$$
with elements being matrices of size $3\times3$, explicitly given  below ($\gamma^{-1}=\sqrt{1-\sp{V}{V}}$, $\,\chi^{-1}=1-\sp{N}{V}$, $\zeta=\gamma\chi{\sp{V}{\Omega}}$),
\bigskip
\begin{small}
\begin{eqnarray*}
\fl
{}\ll_{\vvt}{{}}={{}}{}\tC\gamma\unit{}{}+
{{}}{\gamma}^3\left(F-\pp\left({F_{,\,\pp}}+\pp{F_{,\,\pp\pp}}\right)\right)\vvt{}
-\gamma{\chi}^2{{}}{{F_{,\,\pp\pp}}}\oot{}
{}\\ \fl \phantom{\ll_{\vvt}}-
{\gamma^{-1}}{\chi}^2\left(\pp\br{2{F_{,\,\pp}}+
{\pp}{F_{,\,\pp\pp}}}+2\,\qq\br{3{F_{,\,\qq}}+
2\left(\pp{F_{,\,\pp\qq}}+\qq{F_{,\,\qq\qq}}\right)}\right)
\nnt{}{}\\ \fl \phantom{\ll_{\vvt}}
{}{{}}{}-{{}}\gamma\chi\left({\pp}^2{F_{,\,\pp\pp}}+2\,\qq\br{
\pp{F_{,\,\pp\qq}}-{F_{,\,\qq}}}\right)({\nvt{}+\vnt{}})
{}\\ \fl \phantom{\ll_{\vvt}}+{\chi}^2{{}}{\tB}({\nomt{}+\ont{}}){}
{}{{}}{}+{{\gamma}^2\chi{{}}\pp{F_{,\,\pp\pp}}}({\vot{}+\ovt{}})
{},{}\\
\fl
\ll_{\vot}{{}}{}={{}}{}\chi{}{}{{F_{,\,\pp}}}{}\unit{}+
{2{\chi}^3{}{F_{,\,\pp\qq}}}{}\oot{}{}
{}+{{\gamma}^2\chi{}{}}{\pp{F_{,\,\pp\pp}}}\vvt{}{} \\ \fl\phantom{\ll_{\vot}} +  {{\chi}^2 {\gamma{}{}}^{-2}\left(2{\br{{}\pp
+\zeta}}\tA  -{\gamma}^2{}\tB\right)} \nnt{}{}{}\\ \fl\phantom{\ll_{\vot}}
{{}}{}+{{}}{\chi}^2{{}}{\tB}\nvt{}
{}{-2\,{\gamma^{-1}{\chi}^3{}}\tA}\nomt{}
\\ \fl\phantom{\ll_{\vot}}
{}{{}}{}+{{}}{\chi{{}}\left(2\br{{}\pp+\zeta}\br{\pp{F_{,\,\pp\qq}}
-{F_{,\,\qq}}}-{\gamma}^2{}\pp{F_{,\,\pp\pp}}\right)}\vnt{}
{}-{\gamma{\chi}^2{}}{{F_{,\,\pp\pp}}}\ovt{}\\ \fl\phantom{\ll_{\vot}}
{}{{}}{}-{{}}{2\,{\gamma{\chi}^2{}}\left(\pp{F_{,\,\pp\qq}}-
{F_{,\,\qq}}\right)}\vot{}
{}+{{{}\gamma^{-1}}{\chi}^2\left({}{\gamma}^2{F_{,\,\pp\pp}}-
2\,\left({}\pp+\zeta\right){F_{,\,\pp\qq}}\right)}\ont{},
{}{}\\
\fl
\ll_{\oot}{{}}={{}}{}{-2\,{\gamma^{-1}{\chi}^2{}}{F_{,\,\qq}}}\unit{}
{} -{\gamma{\chi}^2{}}{{F_{,\,\pp\pp}}}\vvt
{-4\,{\gamma^{-1}{\chi}^4{}}{F_{,\,\qq\qq}}}\oot \\ \fl\phantom{\ll_{\oot}}   +{{{\gamma}^{-3}}{\chi}^2{}\left({\gamma}^2
{}\left(2\,{{}{}F_{,\,\qq}}+4\br{{}\pp+\zeta}{F_{,\,\pp\qq}}
-{\gamma}^2{}{F_{,\,\pp\pp}}
\right)-4\,{\left({}\pp+\zeta\right)}^2{F_{,\,\qq\qq}}
\right)}\nnt{}\\ \fl\phantom{\ll_{\oot}}
{}{{}}+{{}}{{\gamma^{-1}}{\chi}^2{}\left({\gamma}^2{}{F_{,\,\pp\pp}}-
2\,\left({}\pp+\zeta\right){F_{,\,\pp\qq}}\right)}(\nvt{}+\vnt{})
{}+{2\,{{\chi}^3{}}{F_{,\,\pp\qq}}}(\vot{}+\ovt{})\\ \fl\phantom{\ll_{\oot}}
{{}}+{{}}{2{{\gamma}^{-2}}{\chi}^3{}\left(
2\,\left({}\pp+\zeta\right){F_{,\,\qq\qq}}-
{\gamma}^2{}{F_{,\,\pp\qq}}\right) }(\nomt+\ont)
.\\
\end{eqnarray*}
\end{small}
Above matrices are linear combinations of a $3\times3$ unit matrix $\unit$ and  $9$ elementary matrices $\nnt$, $\nvt$, $\nomt$, $\vnt$, $\vvt$, $\vot$,
$\ont$, $\ovt$, $\oot$ all of size $3\times3$. Sums, products, inverses and transpositions leave this structure invariant.
Up to a factor, all matrices with the same structure can be written as $\unit+\mathbf{NX^T}+\mathbf{VY^T}+\mathbf{\Omega{}Z^T}$, where $\mathbf{X}$, $\mathbf{Y}$, $\mathbf{Z}$ are linear combinations of $\mathbf{N}$, $\mathbf{V}$, $\mathbf{\Omega}$. This trivial observation enables to use the following identity for their determinants   $$\det\br{\unit+\mathbf{NX^T}+\mathbf{VY^T}+\mathbf{\Omega{}Z^T}}=
\left|\begin{array}{ccc}
1+\sp{N}{X}&\sp{N}{Y}&\sp{N}{Z}\\
\sp{V}{X}&1+\sp{V}{Y}&\sp{V}{Z}\\
\sp{\Omega}{X}&\sp{\Omega}{Y}&1+\sp{\Omega}{Z}\\
\end{array}\right|.$$
Now, on account of the
important identity $$\det{\hh}=\det\br{\ll_{\oot}}\det\br{\ll_{\vvt}-
\ll_{\vot}\br{\ll_{\oot}}^{-1}\br{\ll_{\vot}}^T},$$
 the task of computing the Hessian determinant simplifies significantly. The only thing left is to calculate the inverse of $\ll_{\oot}$, which is also easy, since it can be found by solving the linear system of equations $\br{\ll_{\oot}}^{-1}\br{\ll_{\oot}}=\unit$ for $10$ unknown expansion coefficients of $\br{\ll_{\oot}}^{-1}$ in the base of elementary matrices (to do these calculations one may assume that Gramian determinant for scalar products of vectors $\mathbf{N}$, $\mathbf{V}$, $\mathbf{\Omega}$ is nonzero).
 Equipped with this knowledge, one can show, after a lengthy and tedious but straightforward calculation, that
 \begin{small}
\begin{eqnarray*}\fl\phantom{\det{\hh}}
\det{\hh}=-\frac{\left(F-\pp\,F_{,\,\pp}\right)\left({F_{,\,\pp}}^2+2
\,F_{,\,\qq}\br{F-\pp\,F_{,\,\pp}} \right)}{\,{\left(1-{\sp{N}{V}}\right)}^4\,
{\left(1-{\sp{V}{V}}\right)}^2}\times\dots\\ \phantom{\det{\hh}=}
\dots\left(\br{{F_{,\,\pp}}^2+ {F_{,\,\sqrt{\qq}}}^2}F_{,\,\pp\pp}+\left( F - {\pp}{F_{,\,\pp}}  \right)
\left|\frac{\partial\br{F_{,\,\pp},F_{,\,\sqrt{Q}}}}{\partial
\br{\pp,\sqrt{\qq}}}\right|
\right).\end{eqnarray*}
\end{small}\noindent
For comparison, it is interesting to calculate the Jacobian determinant of a mapping  $\br{\pp,\qq}$ $\to$ $\br{PP\br{\pp,\qq},WW\br{\pp,\qq}}$
\begin{small}
\begin{eqnarray*}\fl\phantom{\det{\hh}}
\frac{\left|\frac{\partial\br{PP,WW}}{\partial
\br{\pp,\qq}}\right|}{-2\,m^6\,{\lambda }^2\,\left( {}{F_{,\,\pp}}\,
     \left( {\pp }^2 + \qq  \right)  - \pp\,F  \right)}=\,\left( {{}{F_{,\,\pp}}}^2 +
    2\,{}{F_{,\,\qq}}\,\left( F - {\pp}{F_{,\,\pp}}  \right)
        \right) \times\dots \\ \phantom{\det{\hh}aa}\dots
\left(\br{{F_{,\,\pp}}^2+ {F_{,\,\sqrt{\qq}}}^2}F_{,\,\pp\pp}+\left( F - {\pp}{F_{,\,\pp}}  \right)
\left|\frac{\partial\br{F_{,\,\pp},F_{,\,\sqrt{Q}}}}{\partial\br{\pp,\sqrt{\qq}}}\right|
\right)
.\end{eqnarray*}
\end{small}
The Jacobian is proportional to $\det{\hh}$, at least if ${F_{,\,\pp}}\,
     \left( {\pp }^2 + \qq  \right)  - \pp\,F\ne0$.

\section*{References}


\begin{thebibliography}{6}
\bibitem{bib:astar1} A. Staruszkiewicz \textit{Fundamental Relativistic Rotator}, \textit{Acta Phys. Pol.} \textbf{B} (2008), Vol. 1, No. 1, 109-112
\bibitem{bib:wigner} E.P. Wigner \textit{On unitary representations of the inhomogeneous Lorentz group}, Ann. Math. \textbf{40}, 149 (1939).
\bibitem{bib:segal} S.M. Kuzenko, S.L. Lyakhovich, A.Yu. Segal \textit{A Geometric Model of Arbitrary Spin Massive Particle}, Int. J. Mod. Phys., \textbf{A10}, 1529 (1995).
\bibitem{bib:hanson} A.J. Hanson, T. Regge \textit{The Relativistic Spherical Top}, \textit{Ann. Phys. (N.Y.)}  \textbf{87}, 498 (1974).
\bibitem{bib:schaefer} V. Kassandrov, N. Markova, G. Schaefer, A. Wipf \textit{On the model of a classical relativistic particle of unit mass and spin}, 	 \texttt{arXiv:0902.3688v2 [hep-th]} (to appear in J. Phys. A: Math. Theor)
\bibitem{bib:bratek} {\L}. Bratek \textit{Nonuniqueness of free motion of the fundamental relativistic rotator},	\texttt{arXiv:0902.4189v2 [math-ph]} (version v1 + material concerning spinning particle \cite{bib:segal})

\end{thebibliography}
\end{document}